\documentclass[superscriptaddress,twocolumn,prl]{revtex4-2}

\usepackage{graphicx}
\usepackage{amsmath}
\usepackage{amssymb}
\usepackage{amsthm}

\usepackage[utf8]{inputenc}

\usepackage{notoccite}

\usepackage{xcolor}
\usepackage{titlesec}

\usepackage{float}
	
\usepackage{hyperref}
\usepackage[all]{hypcap}
\hypersetup{bookmarksnumbered=true,
	colorlinks = true,    
	allcolors = blue,            
	breaklinks = true,   
	backref = true               
}

\let\titleoriginal\title           
\renewcommand{\title}[1]{          
    \titleoriginal{#1}
    \newcommand{\thetitle}{#1}        
}

\DeclareMathSymbol{\shortminus}{\mathbin}{AMSa}{"39} 

\newcommand\titlelowercase[1]{\texorpdfstring{\lowercase{#1}}{#1}}

\titleformat{\section}
{\normalfont\fontsize{12}{15}\bfseries}{\thesection}{1em}{}

\begin{document}


\title{Duality and degeneracy lifting in two-dimensional electron liquids on SrTiO$_3$(001)} 

\author{Igor Sokolovi{\'c}}
\affiliation{Institute of Applied Physics, TU Wien, 1040 Vienna, Austria}

\author{Eduardo B. Guedes}
\affiliation{Photon Science Division, Paul Scherrer Institut, CH-5232 Villigen, Switzerland}

\author{Thomas P. van Waas}
 \affiliation{European Theoretical Spectroscopy Facility, Institute of Condensed Matter and Nanosciences, Universit\'e catholique de Louvain, Chemin des \'{E}toiles 8, B-1348 Louvain-la-Neuve, Belgium}
\author{Samuel Ponc\'{e}}
 \affiliation{European Theoretical Spectroscopy Facility, Institute of Condensed Matter and Nanosciences, Universit\'e catholique de Louvain, Chemin des \'{E}toiles 8, B-1348 Louvain-la-Neuve, Belgium}
\affiliation{WEL Research Institute, Avenue Pasteur 6, 1300 Wavre, Belgium}

\author{Craig Polley}
\affiliation{MAX IV Laboratory, Lund University, Lund, Sweden}

\author{Michael Schmid}
\author{Ulrike Diebold}
\affiliation{Institute of Applied Physics, TU Wien, 1040 Vienna, Austria}

\author{Milan Radovi{\'c}}
\affiliation{Photon Science Division, Paul Scherrer Institut, CH-5232 Villigen, Switzerland}

\author{Martin Setv{\'i}n}
\affiliation{Institute of Applied Physics, TU Wien, 1040 Vienna, Austria}
\affiliation{Department of Surface and Plasma Science, Faculty of Mathematics and Physics, Charles University, 180 00 Prague 8, Czech Republic}

\author{J. Hugo Dil}
\affiliation{Institut de Physique, \'{E}cole Polytechnique F\'{e}d\'{e}rale de Lausanne, CH-1015 Lausanne, Switzerland}
\affiliation{Photon Science Division, Paul Scherrer Institut, CH-5232 Villigen, Switzerland}

\begin{abstract}

	Two-dimensional electron liquids (2DELs) have increasing technological relevance for ultrafast electronics and spintronics, yet significant gaps in their fundamental understanding are exemplified on the prototypical SrTiO$_3$. We correlate the exact SrTiO$_3$(001) surface structure with distinct 2DELs through combined microscopic angle-resolved photoemission spectroscopy and non-contact atomic force microscopy on truly bulk-terminated surfaces that alleviate structural uncertainties inherent to this long-studied system.  The SrO termination is shown to develop a 2DEL following the creation of oxygen vacancies, unlike the intrinsically metallic TiO$_2$ termination. Differences in degeneracy of the 2DELs, that share the same band filling and identical band bending, are assigned to polar distortions of the Ti atoms in combination with spin order, supported with the extraction of fundamental electron-phonon coupling strength. These results not only resolve the ambiguities regarding 2DELs on SrTiO$_3$ thus far, but also pave the way to manipulating band filling and spin order in oxide 2DELs in general. 	

\end{abstract}

\maketitle

The extensive pursuit for the next-generation electronics steers towards low-dimensional systems. Here, SrTiO$_3$ --- a model cubic perovskite oxide --- plays a versatile role. Starting as a promising candidate for high-temperature superconductivity~\cite{muller1979srti}, through its potential as a gate dielectric with $\epsilon_r$ in the order of $10^3$~\cite{yang2022epitaxial}, and finally as an insulator that hosts highly mobile two-dimensional electron liquids (2DELs) at its bare surface ~\cite{santander-syro2011two,meevasana2011creation,wang2014anisotropic,king:2014,guedes2020single,guedes2021universal,Rebec:2019,yan2022origin} or when interfaced with other insulators~\cite{thiel2006tunable,okamoto2006lattice,chikina2018orbital}. Since the discovery of a 2DEL at the LaAlO$_3$/SrTiO$_3$ interface~\cite{ohtomo2004high} it was the subject of two decades of research, but the concepts crucial for oxide electronics~\cite{takagi2010emergent} such as the origin, creation, filling, and lifting of the spin degeneracy in these 2DELs remain debated.
We directly elucidate and disentangle these fundamental phenomena by studying SrTiO$_3$ surfaces of unprecedented quality in direct and reciprocal space. 


\begin{figure*}[t]
	\begin{center}
		\includegraphics[width=2.0\columnwidth,clip=true]{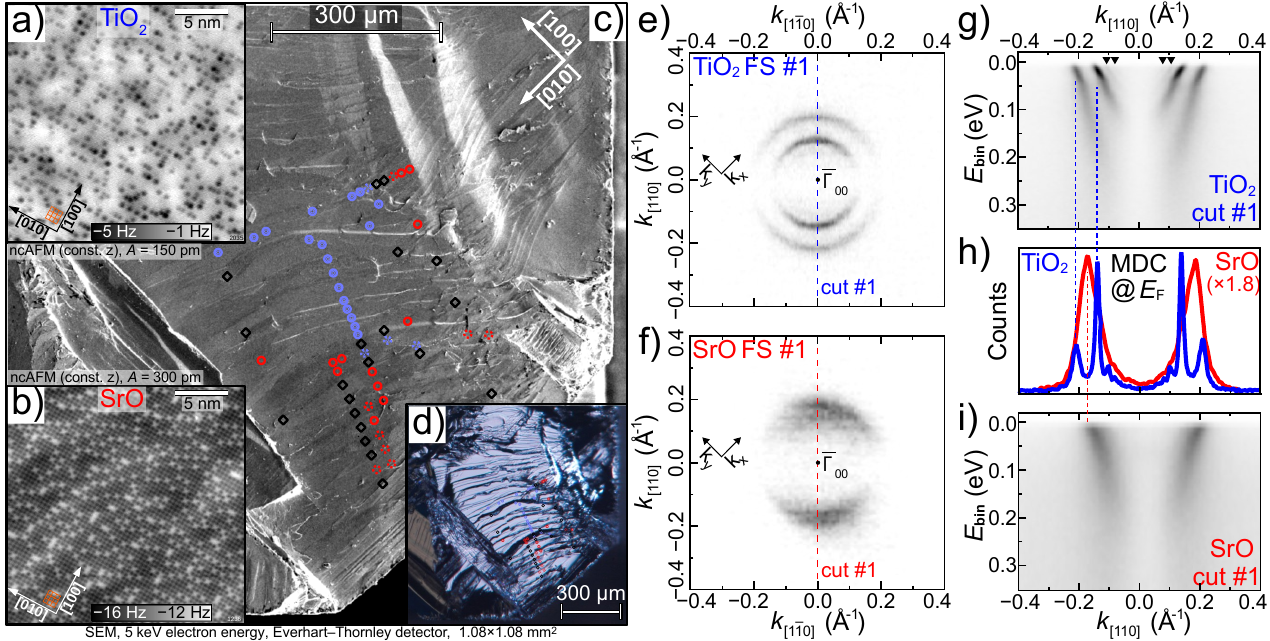}
	\end{center}
	\caption[
	[\textbf{Two distinct 2DELs on a truly bulk-terminated SrTiO$_3$(001)-(1$\times$1) surface.} Atomically resolved ncAFM images obtained on the (a) TiO$_2$-termination with 14\% of Sr adatoms and (b) SrO-termination with 14\% of Sr vacancies. The point defects are intrinsic to a SrTiO$_3$(001) surface cleaved via the strain-induced ferroelectric phase transition. An orange grid indicates the surface unit cells. The counterpiece of the cleaved SrTiO$_3$(001) surface investigated with ARPES displayed in a (c) secondary-electron SEM image and in an (d) optical photograph: Dark SEM contrast corresponds to TiO$_2$- and light gray to SrO-terminated surface regions. Blue markers in (c) and (d) indicate where the measured Fermi surface shows a double ring (e), while red markers show regions with a single ring (f), obtained with  $h\nu$=47 eV (FS \#1). Black diamonds and broken colored circles are positioned where ARPES detected an ill-defined Fermi surface or a Fermi surface that only weakly resembles one of the two well-defined ones, respectively (shown in Fig.~SM4). Marker sizes are to scale with the 10 $\mu$m radius of the beam spot. The corresponding band dispersion map along the $k_{[110]}$ direction (dashed lines, cut \#1) are shown in (g) and (i). MDCs along the dashed lines in (e) and (f) at $E_{\mathrm{F}}$ are shown in (h).]
	{\textbf{Two distinct 2DELs on a truly bulk-terminated SrTiO$_3$(001)-(1$\times$1) surface.} Atomically resolved ncAFM images obtained on the (a) TiO$_2$-termination with 14\% of Sr adatoms and (b) SrO-termination with 14\% of Sr vacancies. The point defects are intrinsic to a SrTiO$_3$(001) surface cleaved via the strain-induced ferroelectric phase transition. An orange grid indicates the surface unit cells. The counterpiece of the cleaved SrTiO$_3$(001) surface investigated with ARPES displayed in a (c) secondary-electron SEM image and in an (d) optical photograph: Dark SEM contrast corresponds to TiO$_2$- and light gray to SrO-terminated surface regions. Blue markers in (c) and (d) indicate where the measured Fermi surface shows a double ring (e), while red markers show regions with a single ring (f), obtained with  $h\nu$=47 eV (FS \#1). Black diamonds and broken colored circles are positioned where ARPES detected an ill-defined Fermi surface or a Fermi surface that only weakly resembles one of the two well-defined ones, respectively (shown in Fig.~SM4~\cite{SOM}). Marker sizes are to scale with the 10 $\mu$m radius of the beam spot. The corresponding band dispersion map along the $k_{[110]}$ direction (dashed lines, cut \#1) are shown in (g) and (i). MDCs along the dashed lines in (e) and (f) at $E_{\mathrm{F}}$ are shown in (h).}
	\label{fig1}
\end{figure*}


\begin{figure*}[t]
	\begin{center}
		\includegraphics[width=2.0\columnwidth,clip=true]{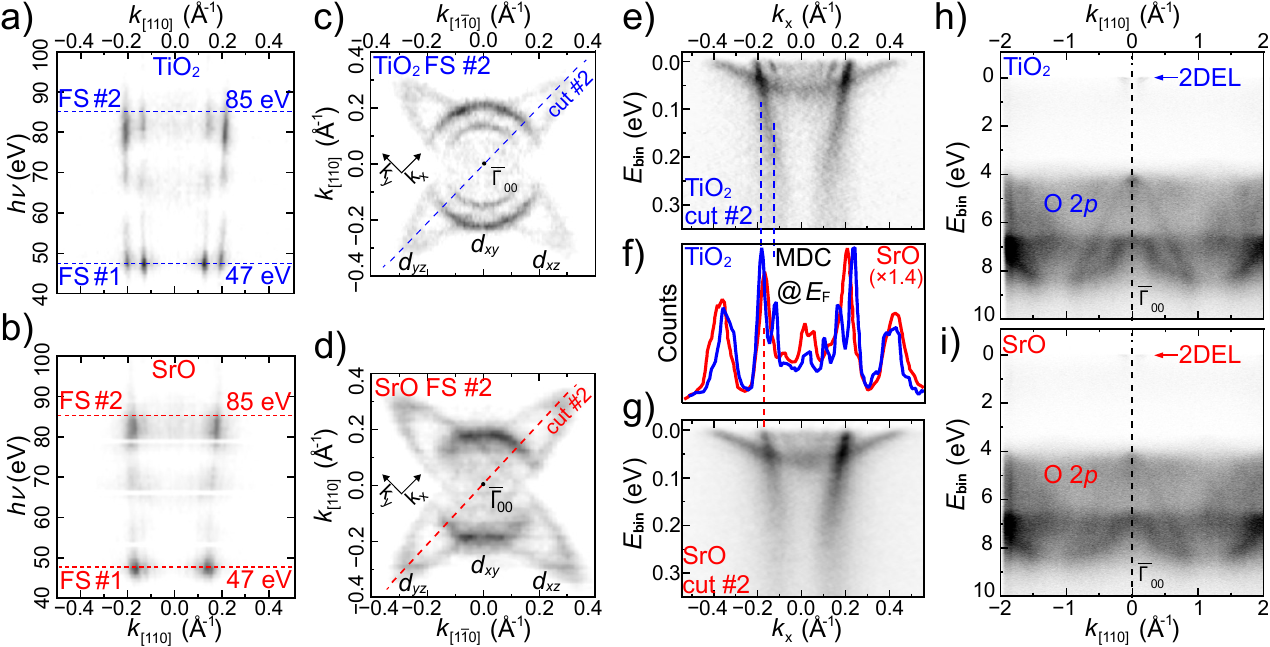}
	\end{center}
	\caption{
		\textbf{Details of the 2DELs on TiO$_2$- and SrO-terminated surfaces.} Photon-energy dependence of the 2DEL formed on the (a) TiO$_2$- and (b) SrO-terminated surfaces. The Fermi surface maps measured with h$\nu$=85~eV (FS \#2) are shown in (c) and (d), and the corresponding band dispersion maps along the $k_x$ direction (cut \#2) in (e) and (g), with a comparison of the MDCs at the Fermi level displayed in (f).  (h,i) O 2$p$ band of the TiO$_2$ and SrO-terminated surface, respecively, measured with $h\nu$=$170$~eV.}\label{fig2}
\end{figure*}

\begin{figure*}[t]
	\begin{center}
		\includegraphics[width=2.0\columnwidth,clip=true]{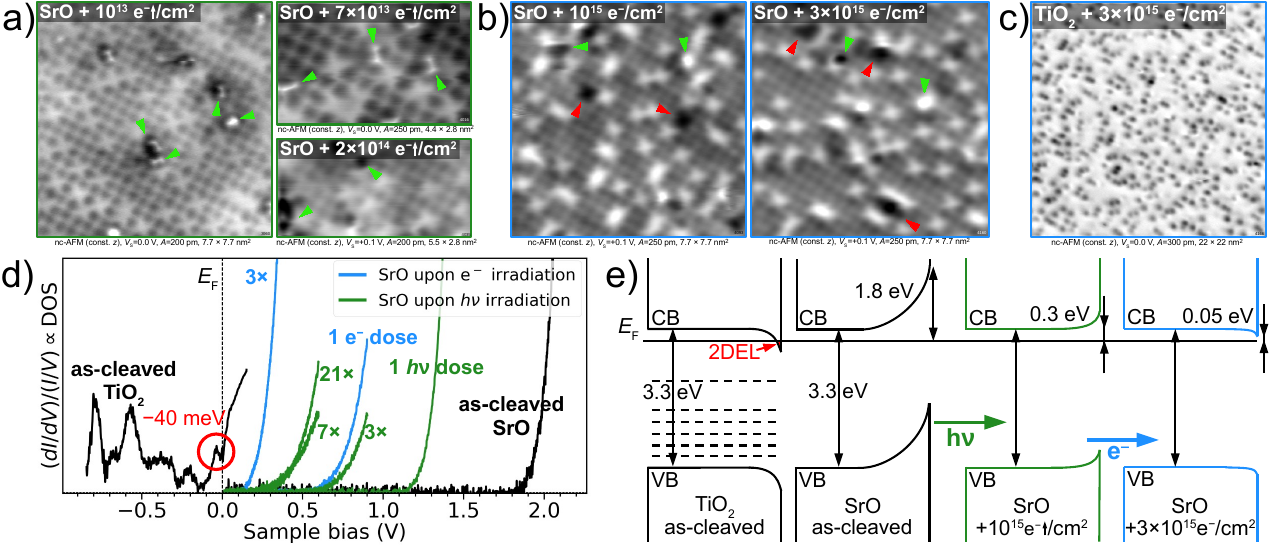}
	\end{center}
	\caption{
		\textbf{2DEL formation mechanism on the SrO-terminated surface emulated with the irradiation of laboratory Al $K\alpha$ X-rays and 47~eV electrons.}
		Atomically resolved ncAFM images obtained on (a) the SrO termination irradiated with several doses of X-rays (measured as photocurrent ``$e^-\uparrow$''; 1 $h\nu$ dose equals $10^{13} e^-\uparrow/\mathrm{cm}^2$), (b) SrO termination and (c) TiO$_2$ termination after two different doses of electron irradiation (1 e$^-$ dose equals $10^{15}e^-/\mathrm{cm}^2$). Green arrows indicate features attributed to adsorbates and red arrows indicate oxygen vacancies that appear exclusively on the SrO termination after irradiation with 47~eV electrons. (d) STS shows the effect of the irradiation on the density of states: TiO$_2$ is unaffected while photoemission and the formation of oxygen vacancies bring the onset of the SrO conduction band closer to the Fermi level.	(e) Illustration of the band structure of the two terminations upon irradiation: TiO$_2$ remains metallic with an intrinsic 2DEL, while the upward band bending on SrO is reduced through photoemission followed by the additional downward band bending induced by the formation of oxygen vacancies.
	}\label{fig3}
\end{figure*}

\begin{figure}[t]
	\begin{center}
		\includegraphics[width=1.0\columnwidth,clip=true]{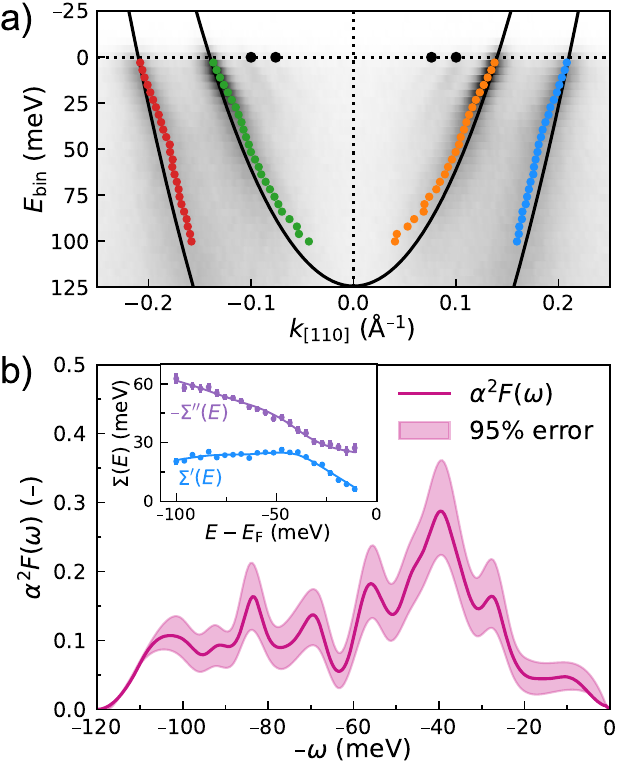}
	\end{center}
	\caption{\textbf{Bare bands, MDC maxima, and the right-hand outer Eliashberg spectral function for the TiO$_2$-terminated surface.} 
		(a) Bare bands (lines) and MDC maxima (colored dots) for the intense bands together with the faint band Fermi wavevectors (black dots).
		(b) The Eliashberg function $\alpha^2F(\omega)$ determined from the right-hand outer real $\Sigma'(E)$ and minus imaginary $\shortminus\Sigma''(E)$ self-energy data, shown in the inset with 95\% confidence intervals and accompanied by continuous lines reconstructed from the obtained $\alpha^2F(\omega)$.
	}
	\label{fig4}
\end{figure}

\vspace*{-0.7cm}
\section{B\titlelowercase{ulk-truncated} S\titlelowercase{r}T\titlelowercase{i}O$_3$(001) surface}
\vspace*{-0.4cm}

Most ambiguities related to the 2DELs on SrTiO$_3$(001) stem from its elusive surface structure, commonly assumed as bulk-truncated yet without real-space confirmation; we demonstrate below the sensitivity of 2DELs towards the exact surface termination. 

With our recently developed cleaving technique that exploits the strain-induced ferroelectric phase transition in SrTiO$_3$~\cite{sokolovic2019incipient}, we routinely create surfaces that are truly bulk-terminated, atomically flat, and well-defined at the atomic level as confirmed by noncontact atomic force microscopy (ncAFM). This is in sharp contrast to differently-prepared surfaces that are either reconstructed, disordered, or contaminated \cite{sokolovic2021quest} due to the tripartite composition sensitive to stoichiometry changes. 

Figure~\hyperref[fig1]{\ref{fig1}} shows ncAFM  and angle-resolved photoemission spectroscopy (ARPES) data on a truly bulk-terminated SrTiO$_3$(001) surface. Our cleaving procedure always produces surfaces with both terminations (TiO$_2$ and SrO), each covered with the same concentration of complementary point defects: TiO$_2$ with 14\% of Sr adatoms (Sr$_{\mathrm{ad}}$) and SrO with 14\% of strontium vacancies (V$_{\mathrm{Sr}}$), resolved by ncAFM in Figs.~\hyperref[fig1]{\ref{fig1}a}~and~\hyperref[fig1]{\ref{fig1}b}, respectively. Their presence is unavoidable as they compensate the polarity induced during successful cleaving~\cite{sokolovic2019incipient,noguera2000polar,goniakowski2007polarity}, but, in turn, guarantee the structure of our well-cleaved surface areas. The O-terminated ncAFM tip~\cite{sokolovic2020resolving} detects Sr$_\mathrm{ad}$ as dark spheres (attraction) and V$_{\mathrm{Sr}}$ as bright ``x''-shaped features (absence of attractive signal). 

At the standard doping level of 0.5~wt\% of Nb, the domains of these two surface terminations are sufficiently large to be individually studied by synchrotron-based ARPES with a beam spot diameter of $\approx 10~\mu\mathrm{m}$. On a mesoscopic scale, the two terminations can be distinguished by scanning electron microscopy (SEM) even after transfer through air: TiO$_2$- and SrO-terminated surface regions are imaged as dark and bright, respectively (details in the Supplemental Material (SM) Sec. SM1~\cite{SOM}). Figure~\hyperref[fig1]{\ref{fig1}c} shows a SEM micrograph of the counterpiece (i.e. the other side after cleavage) of the surface studied under synchrotron light, with the superimposed lateral positions of the ARPES measurements; an optical photograph is shown in Fig.~\hyperref[fig1]{\ref{fig1}d}. With SEM acting as a bridge between synchrotron and ncAFM measurements, we proceed to correlate the 2DEL properties with the surface structure. 

\vspace*{-0.7cm}
\section*{T\titlelowercase{ermination-specific} 2DEL\titlelowercase{s}}
\vspace*{-0.4cm}

The two terminations formally differ in one atomic layer and can hardly be discriminated with x-ray photoelectron spectroscopy (Sec.~SM4~\cite{SOM}), yet each displays a distinct Fermi surface (FS) as witnessed by ARPES (photon energy $h\nu$=$47$~eV) in Figs.~\hyperref[fig1]{\ref{fig1}e,f}. Both are derived from Ti 3$d_{xy}$ orbitals \cite{santander-syro2011two}, but the TiO$_2$ termination is characterized by two sharp rings, while the SrO termination is characterized by a single broader ring. Spectra acquired at different spots on the sample demonstrate that the presented data is representative of the TiO$_2$ and SrO terminations throughout the sample. In contrast, conchoidal surface regions, which lack a well-defined atomic structure \cite{sokolovic2019incipient}, exhibit a variety of less-defined FSs that resemble a combination of the well-defined FSs obtained on the two terminations (Sec.~SM2~\cite{SOM}).  

Band maps measured along the $k_{[110]}$ direction (Figs.~\hyperref[fig1]{\ref{fig1}g,i}) show that these states at the two terminations have a similar dispersion, while the main difference lies in the splitting of the bands. A comparison of the momentum distribution curves (MDCs) at the Fermi level $E_{\mathrm{F}}$ in Fig.~\hyperref[fig1]{\ref{fig1}h} highlights some important differences between the TiO$_2$- and SrO-terminated surfaces. The lines on the former are significantly sharper and the peak maximum of the latter lies between the split bands of the former. This difference cannot be explained by a spectral broadening (Sec.~SM3~\cite{SOM}). Rather, the SrO MDCs also display a shoulder near the inner TiO$_2$ peak, suggesting that the SrO signal is composed of multiple contributions.
Furthermore, the lower intensity on the SrO termination agrees with the localisation of the 2DEL in the subsurface TiO$_2$ plane. 

Figure~\ref{fig2} presents more details of the electronic structure of the 2DELs at both terminations, with the extracted experimental values of $k_{\mathrm{F}}$, carrier densities $n_{2\mathrm{D}}$, and band bottoms $E_{\mathrm{bot}}$ laid out in Tab.~SM1~\cite{SOM}. The absence of dispersion in the photon energy scans (Figs.~\hyperref[fig2]{\ref{fig2}a,b}) confirm that the $d_{xy}$-derived states on both terminations are two-dimensional. Additionally, the resonant enhancements on both terminations occur in the same energy ranges, again indicating their similar orbital composition. 

At $h\nu$=$85$~eV, in addition to Ti 3$d_{xy}$ states of Figs.~\hyperref[fig1]{\ref{fig1}e--i}, the Ti 3$d_{xz}$- and $d_{yz}$-derived states \cite{santander-syro2011two} are visible in the FS maps (Figs.~\hyperref[fig2]{\ref{fig2}c,d}). The Fermi wavevectors $k_{\mathrm{F}}$ and band bottoms of these ``heavy'' bands (Figs.~\hyperref[fig2]{\ref{fig2}e--g}) are \textendash{} within the experimental resolution \textendash{} the same for both terminations (Tab.~SM1~\cite{SOM}). The quality of the in-situ cleaved SrTiO$_3$(001) surface allows us to clearly record the O~$2p$ band on both terminations (Figs.~\hyperref[fig2]{\ref{fig2}h,i}). These measurements were performed with $h\nu$=$170$~eV, allowing the observation of the full first Brillouin zone in the ${<}110{>}$ direction. The similar data on both terminations reiterates that the differences observed in the 2DELs are intrinsic and unrelated to surface quality. 

\vspace*{-0.7cm}
\section{O\titlelowercase{rigin of the} 2DEL\titlelowercase{s}}
\vspace*{-0.4cm}

A 2DEL is known to appear at SrTiO$_3$(001) surfaces after prolonged synchrotron irradiation~\cite{santander-syro2011two,meevasana2011creation,Plumb:2014,Chikina:2019,guedes2020single,Walker:2022}, hinting towards defect-induced population. Both distinct 2DELs on our cleaved SrTiO$_3$(001) surfaces develop much faster: The FS on TiO$_2$-termination becomes clear and intense in a $\lesssim 2$~s time frame (at around 5$\times10^{10}$~$h\nu /\mu$m$^2$), whereas the FS at the SrO termination develops more slowly, and becomes well-defined only after $\approx$30~s (videos in Sec. SM9~\cite{SOM}). This indicates fundamental differences between the two terminations, but also highlights how none or very little defects are necessary for their creation on the TiO$_2$ and SrO terminations, respectively. In comparison, conchoidal surface regions that lack proper atomic order and intrinsic polarization exhibited slowly developing, badly defined FSs (Sec.~SM2~\cite{SOM}). 

In our scanning tunneling spectroscopy (STS) experiments (Fig.~\hyperref[fig3]{\ref{fig3}d}) the TiO$_2$ termination is observed to be metallic immediately after cleaving~\cite{sokolovic2019incipient}. Positive Sr$_{\mathrm{ad}}$ ions compensate the strain-induced polarity during cleaving, yet their presence induces downward band bending on the extrinsically n-doped crystals. 
The presence of distinct in-gap filled states in STS on the TiO$_2$-termination is consistent with the ARPES measurements, but the tip-induced shifting of all energy states in STS and the different tunneling probabilities to distinct $d_{xz/yz}$ and $d_{xy}$ derived states thwart the direct comparison between the two techniques. 

Ab-initio slab calculations show that for the TiO$_2$ termination the $d_{xy}$ states are above the heavy bands in energy~\cite{santander-syro2011two, guedes2021universal}, not at higher binding energy as seen in the ARPES data. This suggests the possibility that the $\lesssim 2$~s delay in detection of a 2DEL on TiO$_2$ is due to the fact that the $d_{xy}$ bands become populated by photo-excited secondary electrons following a small amount of synchrotron irradiation. 
Alternatively, a small amount of adsorbates that trap the charges are quickly removed by the synchrotron light via dissociation/association and desorption~\cite{sokolovic2020resolving}. In either scenario, the creation of defects under synchrotron irradiation is clearly not necessary for a TiO$_2$ termination to host a fully developed 2DEL. 

The SrO termination is wide-gap semiconducting immediately upon cleaving~\cite{sokolovic2019incipient}, promoted by the negative V$_{\mathrm{Sr}}$s that cause an upward band bending (Fig.~\hyperref[fig3]{\ref{fig3}d}). The formation mechanism of a 2DEL, i.e., the semiconductor-to-metal transition on SrO, was investigated separately by progressive irradiation with in-house X-ray and electron sources at $T$=$100$~K, summarized in Fig.~\ref{fig3}. Photoemission by Al $K\alpha$ X-rays (measured as photo-current ``$e^-\uparrow$'') significantly reduces the upward band bending and brings the conduction band onset down by 1.5~eV towards the Fermi level. This downshift saturates with increasing X-ray dose (Fig.~\hyperref[fig3]{\ref{fig3}d}) and, according to STS, is not enough to turn SrO metallic. Ionic defects were not observed with ncAFM following X-ray irradiation, except for the appearance of adsorbates (green arrows in Fig.~\hyperref[fig3]{\ref{fig3}a}). The formation of a 2DEL requires additional downward band bending. Without a synchrotron source, this was achieved by irradiation of the SrO termination with electrons of sufficient energy to create positively charged oxygen vacancies (V$_{\mathrm{O}}$s) via the Knotek-Feibelman  mechanism~\cite{knotek1978ion,dulub2007electron}. The V$_{\mathrm{O}}$s were observed in ncAFM as dark defects in the O sublattice~\cite{setvin2018polarity} only after 47~eV electron irradiation, marked by red arrows in Fig.~\hyperref[fig3]{\ref{fig3}b}. A total of 0.05 monolayers coverage of V$_{\mathrm{O}}$s was sufficient to bring the onset of the conduction band within 50~meV of the Fermi level according to STS (Fig.~\hyperref[fig3]{\ref{fig3}d}). Taking into account tip-induced band bending, this will either be enough or a slightly higher concentration is expected to populate the conduction band. Therefore, in the 30~s of 47~eV synchrotron irradiation, a joint effect of reducing the band bending by photoemission and causing additional downward band bending by the creation of V$_{\mathrm{O}}$s gradually develops a stable 2DEL on the SrO-terminated surface. 

The electronic structure of a metallic TiO$_2$ termination remained unaffected under the same irradiation conditions, and no ionic defects were observed in ncAFM even after the highest dose of 47~eV electrons (Fig.~\hyperref[fig3]{\ref{fig3}c}). In fact, the TiO$_2$ termination with Sr$_{\mathrm{ad}}$s is more prone to adsorption of oxygen to neutralize the Sr ions~\cite{sokolovic2021quest} than the formation of V$_{\mathrm{O}}$s. Therefore, we can safely exclude the relevance of oxygen vacancies for the formation of a 2DEL on the TiO$_2$-terminated surface, while their presence is necessary for a 2DEL on the SrO termination. The presence of two types of defects (V$_\mathrm{Sr}$ and V$_\mathrm{O}$) on the irradiated SrO termination can also explain the larger linewidth and the observed multiple contributions, in contrast to the pristine TiO$_2$ termination. 

\vspace*{-0.7cm}
\section{P\titlelowercase{opulation}}
\vspace*{-0.4cm}

The nearly identical charge density of the 2DEL on both surface terminations, as observed in ARPES, indicates that a mechanism similar to Fermi level pinning at a metal-semiconductor interface is involved in the population of the 2DELs. The observed carrier densities (Tab.~SM1~\cite{SOM}) cannot be rationalized by electron donation from the intrinsic Sr$_\mathrm{ad}$s on the TiO$_2$ termination and the irradiation-induced V$_\mathrm{O}$s on the SrO termination. Real space measurements (Fig.~\hyperref[fig3]{\ref{fig3}}) show that the concentration of these defects is either one or two orders of magnitude too dilute, depending on the termination. Instead, the highly mobile 2DEL electrons stem from the bulk of the sample once the downward band-bending makes the Ti $3d_{xy}$ states energetically available. The V$_\mathrm{O}$s on the SrO-termination hence serve primarily as a band-bending-reversal mechanism, demonstrating a pathway to populating 2DELs with a small amount ($\approx1\%$) of external defects and thus maintaining relatively low defect scattering.  

\vspace*{-0.7cm}
\section{D\titlelowercase{egeneracy lifting}}
\vspace*{-0.4cm}

Correlating distinct 2DELs with their well-defined surface terminations allows us to exclude interpretations of split bands on SrTiO$_3$ surfaces as discrete 2DEL eigenstates in the surface potential wedge \cite{Bahramy:2012,meevasana2011creation,wang2014anisotropic}. Our TiO$_2$ and SrO terminations share the same band bending according to the bottom of the heavy bands (Figs.~\hyperref[fig2]{\ref{fig2}e--g}~and~Figs.~\hyperref[fig3]{\ref{fig3}d,e}) and the Luttinger volume of the 2DEL (Tab.~SM1~\cite{SOM}), stabilized and saturated by the pinned Fermi level. In fact, the purity of the cleaved surfaces and the resulting high quality of the ARPES data allow us to detect additional faint bands on the TiO$_2$ termination with a smaller splitting, clearly visible at small binding energies~(black dots in Fig.~\hyperref[fig4]{\ref{fig4}a}); We assign these states to the next quantum well $N=2$ replicas of the outer-most intense bands. 

The duality of 2DELs' electronic structure in all aspects but the band splitting, calls for an explanation based on lifting of the Kramers degeneracy~\cite{kramers:1930,santander:2014,guedes2020single,Krempasky:2024} by symmetry breaking in the lattice. The displacement of the 2DEL-hosting Ti atoms from an inversion center on the TiO$_2$ termination is promoted by their under-coordination at the interface with vacuum, and the out-of-plane polar distortions inherited from the strain-induced phase transition during cleaving. In analogy with ferroelectric systems~\cite{Krempasky:2018,noel2020non}, opposite displacements are expected on the SrO-terminated regions, but the distortions are in part suppressed by the full coordination of Ti atoms in the O-octahedra, and further rendered incoherent through the introduction of O vacancies required for band bending reversal. Furthermore, transition metal oxides are prone to self-trapping charges in the form of polarons~\cite{franchini2021polarons,reticcioli2019interplay,yim2016engineering} leading to local lattice polarization around Ti$^{3+}$ atoms. Localized polarons can be recognized as distinct in-gap states~\cite{wang2022surface,reticcioli2022competing,ellinger2023small} visible in the STS of the as-cleaved TiO$_2$ termination (states below the Fermi level in Fig.~\hyperref[fig3]{\ref{fig3}d}) and in the  ARPES valence band spectra of both terminations (Fig.~SM6~\cite{SOM}). 

A Rashba-like lifting of the band degeneracy~\cite{rashba1960properties} can emerge from such coherent structural distortions on the TiO$_2$ termination. However, the Rashba effect is expected to affect the $d_{xz,yz}$ orbitals more strongly than the $d_{xy}$ orbitals~\cite{varotto:2022}, because of the absence of a node along the $z$-direction in the latter. Furthermore, our self-energy-based analysis of the bands (Sec.~SM8) indicates not Rashba, but Zeeman-type splitting.

In the absence of an external magnetic field, which would affect both distinct 2DELs, or an intrinsic ferromagnetic order, the Zeeman-type lifting of the 2DEL degeneracy suggests that the band splitting on the TiO$_2$ termination stems from magnetic interactions similar to altermagnets~\cite{vsmejkal2022emerging,Krempasky:2024}. In such systems, lifting of the Kramers degeneracy arises from the in-equivalence of Ti atoms with an unpaired spin in the near surface region and does not require significant spin-orbit coupling, while certain spin configurations still generate spin textures in reciprocal space that observe time-reversal symmetry~\cite{Hellenes:2023arxiv}. More specifically, a non-coplanar spin texture observing combined time reversal and translational symmetry ($T\cdot t$) is expected to be facilitated by slight displacements between neighbouring Ti$^{3+}$ atoms. Along these lines, the SrO surface with a degenerate 2DEL represents an antiferromagnet with the spin sublattices connected by translation symmetry, or lacks the spin order entirely.
Further experiments with spin resolution in real and reciprocal space and advanced theoretical approaches will be needed to refine this scenario. 

\vspace*{-0.7cm}
\section{E\titlelowercase{lectron-phonon coupling}}
\vspace*{-0.4cm}

The strong coupling between electronic and structural degrees of freedom on the TiO$_2$-terminated surface is further supported by the observation of a kink in the $d_{xy}$-derived bands at binding energies in the energy range of the SrTiO$_3$ phonon bandwidth~\cite{vogt:1988}, hinting at sizeable electron-phonon coupling (EPC)~\cite{johnson:2001}.
In Fig.~\hyperref[fig4]{\ref{fig4}a}, we reproduce the data from Fig.~\hyperref[fig1]{\ref{fig1}g} and show the bare dispersions (parameters in Tab.~SM1~\cite{SOM}) and MDC maxima.
Using our recently developed self-consistent extraction method~\cite{SOM}, we determine the Eliashberg spectral function $\alpha^2F(\omega)$~\cite{grimvall:1981} of the outer right-hand branch.
The real and minus imaginary parts of the complex electron self-energy $\Sigma(E)=\Sigma'(E)+\mathrm{i}\Sigma''(E)$~\cite{damascelli:2004} are shown in the inset of Fig.~\hyperref[fig4]{\ref{fig4}b}, whereas the extracted $\alpha^2F(\omega)$ is presented in the main figure. The latter shows large spectral weight at $\omega=56$~meV, $\omega=69$~meV, and $\omega = 84$~meV, likely related to the respective LO$_3$, TO$_4$, and LO$_4$ phonon modes~\cite{vogt:1988}, which display moderate to strong EPC in undoped bulk SrTiO$_3$ at similar frequencies~\cite{zhou:2018a}.
Notably, due to charges accumulated at the surface in the form of the 2DEL and in-gap states, our LO$_4$ mode is red-shifted by approximately 8 meV with respect to first-principles calculations of undoped bulk SrTiO$_3$~\cite{zhou:2018a,cancellieri:2016}. The intensity and width of the large spectral weight near $\omega=40$~meV suggests a mixed coupling. Whereas the rest of the modes are well reproduced for the outer left-hand branch (Sec. SM7~\cite{SOM}), this mixture is shifted, hinting at possible anisotropies in the EPC.

Thus, the exceptional data quality for the TiO$_2$-terminated surface has allowed us to identify the dominant EPC modes and extract the Eliashberg function and total EPC parameter $\lambda \equiv - \partial \Sigma'(E)/\partial E|_{E=E_{\mathrm{F}}}=0.63$, which allow for evaluating theories of phonon-mediated superconductivity in SrTiO$_3$~\cite{gastiasoro:2020a}.

The combination of spin order and the strong coupling between electronic and lattice degrees of freedom opens up the thrilling prospect of coherently controlling the spin texture, and thus also the spin-to-charge conversion~\cite{noel2020non}, via ultrashort laser pulses.

\section{C\titlelowercase{onclusions}}
\vspace*{-0.4cm}

Our results highlight that a precise knowledge and control of the surface structure is crucial to understand the fundamental mechanisms and engineer the properties of 2DELs on oxide surfaces. We shine new light on 2DELs on SrTiO$_3$(001) by studying and manipulating truly bulk-terminated surfaces with unambiguous surface structure. Two well-defined terminations, differing in a single atomic layer only, host two distinct 2DELs, while the atomically ill-defined surface areas (conchoidal) can host a myriad of various 2DELs of lesser quality. The distinct 2DELs on both terminations are populated by charges from the bulk of the material, which is found to be self limiting with identical band filling. Irrespective of these similarities, the Kramers degeneracy is lifted on the TiO$_2$, and preserved on the SrO termination, indicating a structural cause and the connection between polar distortions and non-relativistic spin order. The presence of strong electron-phonon coupling makes these surfaces a promising venue to explore such coupling mechanisms on ultrashort time scales and drive spin phenomena by external stimuli.   

\vspace*{-0.7cm}
\section{A\titlelowercase{cknowledgments}}
\vspace*{-0.4cm}
I.S., M.Sc., U.D., and M.Se. acknowledge support by the Austrian Science Fund (FWF) projects Solids4Fun (F-1243) and SuPer (P32148-N36). U.D. also acknowledges support by the FWF SFB project TACO (Grant-DOI:10.55776/F81); for open access purposes the author has applied a CC BY public copyright license to any author accepted manuscript version arising from this submission. T.P. v.W. is a Research Fellow of the F.R.S.-FNRS. S.P. acknowledges financial support from the F.R.S.-FNRS (Belgium).  M.R. and E.B.G  were supported by SNSF Research Grant 200021\_182695. M.Se. also acknowledges support from the Czech Science Foundation GACR 20-21727X and GAUK Primus/20/SCI/009. 
We acknowledge MAX IV Laboratory for time on the Bloch Beamline under Proposal 20210244. Research conducted at MAX IV, a Swedish national user facility, is supported by the Swedish Research council under contract 2018-07152, the Swedish Governmental Agency for Innovation Systems under contract 2018-04969, and Formas under contract 2019-02496.

\vspace*{-0.7cm}
\section{M\titlelowercase{ethods}}
\vspace*{-0.4cm}
\textbf{SrTiO$_3$ single crystals} with 0.5~wt\% of Nb$_2$O$_5$ doping (0.7~at\% Nb at the B sites of SrBO$_3$) were used in this study. The custom shaped 3.5$\times$2$\times$7~mm$^3$ SrTiO$_3$ samples were purchased from MaTeck GmbH. Pristine bulk-terminated strontium titanate SrTiO$_3$(001) surfaces were achieved by cleaving SrTiO$_3$ single crystals at room temperature, through a procedure described elsewhere~\cite{sokolovic2019incipient,sokolovic2021quest}. Cleaving devices made out of stainless steel were thoroughly cleaned ex situ before each cleaving, and were not degassed in vacuum since the investigated SrTiO$_3$(001) surfaces were not exposed to temperatures higher than room temperature. All surfaces shown in the main text were cleaved in situ: samples studied with ncAFM were cleaved inside an UHV chamber with a base pressure lower than 1$\times$10$^{\shortminus10}$~mbar, and samples studied with ARPES were cleaved inside a baked UHV loadlock with the pressure lower than 1$\times$10$^{\shortminus8}$~mbar. After cleaving, the samples were introduced to the measurement chambers with lower pressure as soon as possible, within several minutes.

\textbf{ARPES measurements} were performed at MAX IV synchrotron facility in Lund, Sweden, at the Bloch beamline. Cleaved samples were held at $T=21.5$~K during measurements in an UHV chamber with a base pressure of $1\times10^{\shortminus10}$~mbar. All ARPES measurements were performed with linear vertical, i.e. s-polarized light and the slit of the Scienta DA30 electron analyzer was perpendicular to the scattering plane, and parallel to the $[110]$ orientation of the crystal. The experiments were reproduced on four samples during two experimental runs, with up to 100 measurement sites per sample.

\textbf{Atomically resolved ncAFM measurements} were performed with a low-temperature Omicron qPlus STM/AFM head, located in UHV with base pressure below 10$^{\shortminus11}$~mbar. Bias sweeps used for recording STS measurements and Kelvin parabolas were performed in the same measurement head, by applying bias to the sample. The sample bias was modulated with a frequency of 123~Hz and an amplitude of 10~mV during STS measurements, and the current signal was differentiated using a Z{\"u}rich Instruments lock-in amplifier. All STM/ncAFM measurements were performed close to the LHe temperature, i.e., at $T\approx5$~K. Electrochemically etched tungsten tips cleaned in situ by self-sputtering with Ar$^+$ ions~\cite{setvin2012ultrasharp} were used, after thorough functionalization on a Cu(110) single crystal surface. All images in the main text were achieved by a tip that was additionally functionalized with an O atom at the tip apex~\cite{sokolovic2020resolving}. Custom-design high-quality-factor ($\approx$50000) qPlus tuning forks with a separate wire for the tunneling current were used~\cite{giessibl2013sensor}. The cantilever deflection was measured using a cryogenic differential preamplifier~\cite{huber2017low}. A bungee-cord suspension system was employed for removing mechanical noise during measurements~\cite{schmid2019device}.

\textbf{X-ray and electron irradiation} were both performed at a sample held in an electrically grounded manipulator cooled to $T=100$~K, in an UHV chamber with a base pressure below 1$\times$10$^{\shortminus10}$~mbar. The irradiated surfaces were subsequently studied after their reintroduction to the ncAFM measurement head. The samples were not warmed up following the irradiation at $T=100$~K as they were transferred to the measurement head using a pre-cooled wobblestick at a temperature lower than $T=100$~K. Al $K\alpha$ X-rays were generated from a dual-anode X-ray source (SPECS XR50, Mg/Al~$K\alpha$, operated with 400 W power at a distance of $\approx$10\,mm to the sample). Low-energy electrons were emitted from the filament of the low-energy electron diffraction (LEED) setup and were focused on the sample through integrated electron optics. Electrons with 47~eV were chosen in order to excite the Knoteck-Feibelman mechanism \cite{knotek1978ion}. The spot size for X-ray irradiation was $\approx 1$~cm$^2$, while the electron irradiation was focused to $\approx 1$~mm$^2$. Uniform irradiation of the entire sample was achieved by rastering the sample position below a fixed X-ray or electron source. The X-ray and electron doses were varied by changing the irradiation time while all the other parameters were kept fixed. The X-ray and electron count indicated in Fig.~3 were converted from the measured sample current during irradiation, adjusted for the spot size and the duration of irradiation.

\textbf{SEM images} were acquired with a FEI Quanta 200F measurement setup, with a nominal vacuum of $1\times10^{\shortminus5}$~mbar. The working distance was $\approx 8$~mm. Secondary electron were detected with an Everhart–Thornley detector. All surfaces imaged with SEM were acquired under same conditions. An electron energy of 5~kV was used. SEM imaging was performed at the USTEM facility of the TU Wien.

\textbf{Optical photographs} were taken using an Olympus SZX12 microscope equipped with a DF PLAPO 1X PF lens, and an attached Olympus E-330 camera with a 14-45~mm lens. 

\textbf{Data} acquired with ncAFM was processed for drift correction, background subtraction, and noise reduction using ImageJ software (imagej.net) augmented with procedures developed by Michael Schmid. ARPES data was analyzed and processed using \textit{pesto} Python library (pesto.readthedocs.io) developed by Craig Polley, in combination with IGOR Pro software (wavemetrics.com). Self-energy extraction was performed through procedures developed by Thomas P. van Waas and Samuel Ponc\'{e} laid out in the Supplemental Material~\cite{SOM} and prepared for a separate publication. All raw data is available from authors upon request.

\bibliography{Bibliography_STO_2DELs}
\pagebreak{}



\end{document}